\documentclass{PoS}

\usepackage[stable]{footmisc}
\newcommand{\be}{\begin{eqnarray}}
\newcommand{\ee}{\end{eqnarray}}
\newcommand{\bea}{\begin{eqnarray}}

\newcommand{\eea}{\end{eqnarray}}

\newcommand{\ms}[1]{\textrm{\tiny $#1$}}

\newcommand{\LO}{\ms{(0)}}
\newcommand{\FO}{\ms{(1)}}

\title{Canonical charges and asymptotic symmetries in four dimensional conformal gravity}

\ShortTitle{Canonical charges and asymptotic symmetries in four dimensional conformal gravity}

\author{\speaker{I. Lovrekovic}\\
         Technische Universit\"at Wien, Institute for Theoretical Physics\\
        E-mail: \email{lovrekovic@hep.itp.tuwien.ac.at}}

\abstract{We calculate canonical charges in four dimensional conformal gravity using the generalised boundary conditions  presented in \cite{Grumiller:2013mxa}. We show that the charges are finite and conserved. The asymptotic symmetry algebras generated by these charges are classified according to the conditions imposed on the first subleading term in the generalised Fefferman-Graham expansion.}

\FullConference{Proceedings of the Corfu Summer Institute 2014 "School and Workshops on Elementary Particle Physics and Gravity",\\
		3-21 September 2014\\
		Corfu, Greece }

\begin{document}


\section{Content}

The purpose of this talk is to motivate the study of conformal gravity (CG) via canonical analysis, which is done in chapters two and three. It introduces the mathematical tools used in canonical analysis such as integrability conditions and the Castellani algorithm \cite{Castellani:1981us}, and gives reference to the literature where different procedures, calculation of finiteness and time conservation, are performed, on this we focus in chapters four and five. The described tools are applied to the theory of conformal gravity in chapters six and seven. The canonical charges define the asymptotic symmetry algebra (ASA) that preserves the boundary conditions of the theory, described in chapter eight.  As an example of the ASA that is obtained in CG, we consider the ASA for boundary conditions obtained by analysis of one of the subalgebras of conformal algebra, which is our new result presented here, and the boundary conditions motivated by the Mannheim-Kazanas-Riegert solution of CG \cite{Riegert:1984zz},\cite{Mannheim:1988dj}, shown in chapter nine. In chapter ten we give conclusion of the talk and an outlook of possible further research and applications of the analysis.

\section{Motivation}

Conformal gravity has  recently gained more attention 
after being studied  by Maldacena \cite{Maldacena:2011mk}, who proved that one can impose suitable boundary conditions reducing the solution of CG to that of  Einstein gravity (EG). It was also shown to arise from the twistor string theory and as a boundary counterterm in five dimensional Einstein gravity \cite{Berkovits:2004jj,Liu:1998bu,Balasubramanian:2000pq}. 
In the series of articles by t'Hooft, it was studied as a theory that can possibly give better insight into physics at the Planck scale \cite{Hooft:2010nc,Hooft:2010ac,Hooft:2014daa}.
Interesting phenomenological applications were described by Mannheim, who described the galactic rotation curves within a CG framework without addition of dark matter \cite{Mannheim:1988dj,Mannheim:2010xw,Mannheim:2012qw,Mannheim:2011ds,Mannheim:2006rd}. A toy model for galactic rotational curves without addition of the dark matter was as well studied in  \cite{Grumiller:2010bz}.

Recent work showed that conformal gravity admits more general boundary conditions than those studied by Starobinsky in \cite{Starobinsky:1982mr}. Using the more general boundary conditions yields two response functions analogous to the Brown-York stress energy tensor. One of them can be called precisely Brown-York stress energy tensor since it is sourced by the leading term in the asymptotic expansion of the metric while another one is sourced by the first term in the expansion and is called partially massless response \cite{Grumiller:2013mxa}. Partially massless response in CG was also studied in \cite{Maldacena:2011mk,Deser:2012qg,Deser:2013uy}.

 
 We obtain an interesting solution of the CG equations of motion. It respects the generalised Starobinsky boundary conditions and realises  the asymptotic symmetry algebra which is one of the subalgebras of conformal algebra. The solution  can be conformally rescaled to a Ricci flat solution, studied in \cite{Caldarelli:2013aaa} and \cite{Caldarelli:2012hy}, where it was shown that the holography for the asymptotically flat spacetimes  implies that the Ricci  flat spacetimes that appear in the  AdS/Ricci-flat correspondence inherit holographic properties  from the corresponding AdS spacetime. They have a generalised conformal structure  referred to as "hidden-conformal invariance".
Ricci flat metrics have also been analysed in the discussion of the deformations of supertwistor and ambitwistor spaces in relation with conformal and superconformal symmetries\cite{Lindstrom:2005uh}.



\section{Conformal gravity}

Conformal gravity is a theory of gravity that is invariant under local Weyl rescalings of the metric $g_{\mu\nu}$ on a manifold $\mathcal{M}$

\begin{eqnarray} g_{\mu\nu}\rightarrow\tilde{g}_{\mu\nu}=e^{2\omega}g_{\mu\nu}\label{eq1}.\end{eqnarray}

 Where $\omega$ is Weyl factor which can depend on all the coordinates on $\mathcal{M}$. 
The action of conformal gravity 

\begin{eqnarray}I_{CG}=\alpha_{CG}\int d^4x \sqrt{\left| g\right|}C^{\lambda}{}_{\mu\sigma\nu}C_{\lambda}{}^{\mu\sigma\nu}\label{eq2}\end{eqnarray}
is given by the integral of the square of the Weyl tensor, which is the completely traceless part of the Riemann tensor. The equations of motion 

\begin{equation}
B_{\mu\nu}\equiv\left(\nabla^{\sigma}\nabla_{\lambda}+\frac{1}{2}R^{\sigma}{}_{\lambda}\right)C^{\lambda}{}_{\mu\sigma\nu}=0
\end{equation}
that follow from this action imply the vanishing of the Bach tensor $B_{\mu\nu}$.

\section{Hamiltonian analysis}

In simple dynamical theories momentum variables can be inverted so that velocities can be expressed in terms of the coordinates and the momenta. However that is not true in general  theories. In order to allow for  momenta $p$ to depend on the  velocity we have to introduce constraints that connect coordinates $q$ and momenta $p$
\begin{equation}\phi_{m}(q,p)=0\label{eq3}.\end{equation} Where $\phi_m$ are called primary constraints  and $m=1,...,k$  is number of the constraints \cite{Blagojevic:2002du}. 
If we define the phase space on which $p$ and $q$ are not constrained as $\Gamma$, then the subspace $\Gamma_{1}$ defined by the constraints (\ref{eq3}) is the constrained phase space. The function  \begin{equation}F(p,q)\vert_{\Gamma_1}=0\label{eqn}\end{equation} which vanishes on the constrained phase space is said to vanish \emph{weakly} or to satisfy "weak equality". The function $F$ satisfies "strong equality" when in addition to (\ref{eqn}) its partial derivatives with respect to coordinates $\frac{\partial F}{\partial q}\vert_{\Gamma_1}=0$ and momenta $\frac{\partial F}{\partial p}\vert_{\Gamma_1}=0$ also vanish on the constrained phase space. The weak equality we denote with "$\approx$" and strong equality with "$=$".
Varying the function $F$ on the constrained surface we find it to be equal to zero

\begin{equation}\delta F\vert_{\Gamma_1}=\left(\frac{\partial F}{\partial q_{a}}\delta q_{a}+\frac{\partial F}{\partial p_{a}}\delta p_a\right)\vert_{\Gamma_1}=0.\label{varf} \end{equation}
In which the variations of momenta and coordinates satisfy the $k$ constraints (\ref{eq3}). Therefore we can write
\begin{equation}
\frac{\partial\phi_m }{\partial q_a}\delta q_a + \frac{\partial \phi _m}{\partial p_a}\delta p_a\approx 0.\label{varcons}
\end{equation}

Using the general method of computation of variations with constraints (Ter Haar 1971)
 we find that the terms next to $\delta q_a$ and $\delta p_a$ are equal up to an arbitrary multiplier $\lambda^m$. Inserting that multiplier in the expression (\ref{varcons}) and combining with (\ref{varf}) we obtain 
\begin{equation}
\frac{\partial}{\partial q_a}(F-\lambda^m\phi_m)\approx 0
\end{equation}
and an analogous equation for $\frac{\partial}{\partial p_a}.$
We can conclude that the Hamiltonian which we define as canonical Hamiltonian, can be expressed independently on the velocities, only in terms of $q$s and $p$, i.e. it is valid only on the constraint surface. 
Therefore we define the total Hamiltonian $H_{T}$ which differs from the canonical Hamiltonian by terms proportional to the constraints
\begin{equation} H_T=H_c+\lambda^m \phi_m\label{ht} \end{equation}
where  $H_c$ is canonical Hamiltonian (Legendre transform). Varying  (\ref{ht}) with respect to $\lambda, p, q$ we obtain the constraints and the Hamiltonian equations of motion involving arbitrary multipliers.


Hamiltonian equations of motion can be nicely written in terms of Poisson brackets. The equation of motion $\dot{g}(q,p)$ of an arbitrary dynamical quantity $g(q,p)$  is then given by
\begin{equation}\dot{g}=\{g,H_c \} +u^m \{ g,\phi_m\}\approx \{g,H_T \}  \label{cond} \end{equation}
where $u^m$ are arbitrary multipliers and where it was taken into consideration that the equations are evaluated on shell which means that the quantities that satisfy weak equality are zero. 
In the same sense, the equation of motion of a constraint $\phi_m$ is 

\begin{equation} \dot{\phi}_m= \{\phi_m,H_c \} +u^n\{\phi_m,\phi_n \}.  \label{eomphi} \end{equation}
Consistency of the theory demands that the primary constraints are conserved during the time evolution.  This leads to three possible consistency conditions: 
\begin{itemize}
\item the equality (\ref{eomphi}) can be trivially satisfied, $0=0$
\item the equality (\ref{eomphi}) can lead to equation determining the Legendre multipliers in terms of $p$s and $q$s.
\item the equality (\ref{eomphi}) can lead to an expression that does not contain any of the multipliers. In that case we obtain new \emph{secondary constraints} that define a constraint surface $\Gamma_2\subseteq\Gamma_1$.
\end{itemize}

 If we denote the full set of constraints (primary and secondary) with $\varphi$, the equation of motion for $\varphi$

\begin{eqnarray} \dot{\varphi}_s= \{\varphi_s,H_c \} +u^m\{\varphi_s,\phi_m \}   \label{eomchi} \end{eqnarray}
 with $s=1,...,N$ (N= number of primary and secondary constraints), has a  general solution $u^m=U^m+v^aV_a{}^m$ where
 $V_a{}^m$ are independent solutions of homogeneous equation, $v^a=v^a(t)$ are arbitrary coefficients, the index $a$ runs over all these solutions and $U^m$ are particular solutions of the inhomogeneous equations. Using this general solution, and definition $V_a{}^m\phi_m=\phi_a$, we can see from the total Hamiltonian

\begin{equation}
H_T=H_c+U^m\phi_m+v^aV_a{}^m\phi_m=H_c+U^m\phi_m+v^a\phi_a. \label{ht1}
\end{equation}
 that after all the consistency requirements  are satisfied there are arbitrary functions of time left in the equation. That means that the dynamical variables at some future instant of time are not uniquely determined  by their initial values.

Let us introduce the concepts of the first and second class constraint. If the dynamical variable R(q,p) has a weakly vanishing Poisson bracket with the all the constraints in the theory it is said to be first class. Otherwise it belongs to the second class.  $H_c+u^n\phi_n$ and $\phi^a$ are  first class constraints which we can see  from (\ref{ht1}).%



Due to the arbitrary functions of time left in the equation (\ref{eomchi}), $p(t)$ and $q(t)$ can not be uniquely determined from their initial values. We can show this by considering the general dynamical variable $g(t)$ at $t=0$ and the way it changes in the time interval $\delta t$. Initial values $(q(0),p(0))$ determine initial value $g(0)$.  The value of the function $g(t)$ at time $\delta t$ is calculated from
\begin{eqnarray} 
g(\delta(t))&=&g(0)+\delta t \dot{g} \\
&=&g(0)+\delta t \left( \{g,H'\}+v^a\{g,\phi_a\} \right).
\end{eqnarray}

The arbitrary coefficients $v^a(t)$ allow for different values of these coefficients and give different values of $g(\delta t)$ which is $\Delta g(\delta t)=\delta t(v_2^a-v_1^a) \{g,\phi_a\}$.
Since we know that physical states $g_1(\delta t)$ and $g_2 (\delta_t)$ on the constraint surface are independent of arbitrary multipliers, $g(\delta t)$ must be unphysical.  While H' is a sum of canonical Hamiltonian and $U^m\phi_m$.


Since the number of arbitrary functions $v^a(t)$ is equal to the number of first class constraints $\phi^a$, that means that primary first class constraints (PFCs) generate unphysical transformations of dynamical variables, i.e. gauge transformations \cite{Blagojevic:2002du}. 

\section{Castellani algorithm}

Let us assume that we have a total Hamiltonian (\ref{ht}), complete set of constraints $\phi_b\approx0$, a trajectory $T_1(t)=(q(t),p(t))$ with given initial conditions on the constraint surface $\Gamma_2$ and that we have calculated the functions $v^a(t)$. The  equations of motion are then
\begin{center}
\begin{eqnarray}
\dot{q}_i=\frac{\partial H'}{\partial p_i}+v^a \frac{\partial \phi_a}{\partial p_i} \\
-\dot{p}_i=\frac{\partial H'}{\partial q_i}+v^a \frac{\partial \phi_a}{\partial q_i} \\
\phi_b(q,p)=0 \label{hteom}
\end{eqnarray} 
\end{center}
where $\phi_b$, complete set of constraints, are all the primary and secondary constraints and $b$ their total number. 
Since we noticed that all the primary first class constraints generate gauge transformations we call\newline
$\phi_a$  generator of the transformations, and $\delta v^a$, an infinitesimal time dependent parameter. 
The variation of the dynamical variable  F, $\delta F=\delta v^a\{F,\phi_a\}$ by an arbitrary infinitesimal parameter $\epsilon(t)$ becomes 
\begin{eqnarray}
\delta q_i&=&\epsilon(t)\{q^i,G\}=\epsilon(t)\frac{\partial G}{\partial p_i}\\
\delta p_i&=&\epsilon(t)\{p^i,G\}=-\epsilon(t)\frac{\partial G}{\partial q_i}. \label{varpq}
\end{eqnarray}
 where $G$ is  the generator of this transformation. 
 
Variation of (\ref{hteom}) of with respect to $v^a(t)$ and differentiation of (\ref{varpq}) with respect to $t$  leads to a system of equations from which we obtain the relation 
\begin{equation}
\{F,\dot{\epsilon}G+\epsilon\{G,H_T\}-\phi_a\delta v^a\}\approx 0 \label{trivgen}
\end{equation}
for an arbitrary function $F$ on the surface $\Gamma_1$.  Therefore, we can conclude that  $\dot{\epsilon}G+\epsilon\{G,H_T\}-\phi_a\delta v^a$ is a trivial generator. In general, the generator  $G$ takes the form 
\begin{equation}
G=\epsilon^{(k)}(t)G_k +\epsilon^{(k-1)}(t) G_{k-1}+...+\epsilon G_0\label{defg}
\end{equation}
with $\epsilon^{(k)}=\frac{d^k\epsilon}{dt^k}$. The relation (\ref{trivgen}) is solvable via the algorithm

\begin{eqnarray}
G_k=&PFC\\
G_{k-1}+\{G_k,H\}=&PFC \\
.& \\
.&\\
.&\\
G_1+\{G_2,H\}=&PFC \\
G_0+\{G_1,H\}=&PFC \\
\{G_0,H\}=&PFC 
\end{eqnarray}
developed by Castellani \cite{Castellani:1981us}.

\section{Hamiltonian analysis of conformal gravity}
To apply the above procedure on CG \cite{Kluson:2013hza,Irakleidou:2014vla}, we introduce the first order Arnowitt-Deser-Misner formalism \cite{Arnowitt:1962hi}. Higher derivative theories  in the Hamiltonian formulation were studied in  \cite{Boulware:1983yj,Buchbinder:1987vp,Deruelle:2009zk}. 
It was shown in \cite{Grumiller:2013mxa} that for the  variational principle of the action to be well defined there are no boundary counter terms necessary. 


 A  globally hyperbolic space-time manifold $\mathcal{M}$ admits a foliation into a family of nonintersecting space like Cauchy surfaces $\Sigma_t$. The metric defined on  $\mathcal{M}$ induces metric $\gamma_{\mu\nu}$ on the hypsersurface $\Sigma_t$. We use the indices $\mu,\nu,...$ for the quantities and summation on the four dimensional manifold, and we will use latin indices $a,b,c..$ for the quantities on the three dimensional hyper surface. If we define $n_{\mu}$ as timelike, future directed unit normal to $\Sigma_t$,  its metric can be written as $\gamma_{\mu\nu}=g_{\mu\nu}+n_{\mu} n_{\nu}$  with $n_{\mu}n^{\nu}=-1$.  $\gamma^{\mu}_{\nu}$ simultaneously acts as a projector on $\Sigma_t$ and decomposes tensors in $\mathcal{M}$. 
In order  to introduce  explicit basis, we define timelike vector  $t^{\mu}$ such that $t^{\mu}\partial_{\mu}t=1$, where $\partial_{\mu}$ is the partial derivative on the manifold $\mathcal{M}$. $t^{\mu}$ is in ADM variables decomposed in terms of lapse $N$ and shift $N^a$  as $t^{\mu}=N n^{\mu}+N^{\mu}$ with $N=-n_{\mu}t^{\mu}$ and $N^{\mu}=h^{\mu}{}_{\nu}t^{\nu}$.   $N,N^a$ and $\gamma_{ab}$ define the geometry of the space-time.  Taking $t$ as a time coordinate, and introducing arbitrary coordinate system $x^a, a=1,2,3$ on $\Sigma_t$, we find $n_{\mu}=-N\partial_{\mu}t=(-N,0,0,0)$ and $n^{\mu}=(\frac{1}{N},-\frac{N^{a}}{N})$.
The line element now reads
\begin{equation}
ds^2=-N^2dt^2+\gamma_{ab}(dx^a+N^adt)(dx^b+N^bdt).
\end{equation}

To perform the canonical analysis we start by rewriting the CG action (\ref{eq2}) in terms of that decomposition 
\begin{equation}
S=dt\int_{\Sigma_t}d^3x N \sqrt{\gamma}(-2 \alpha_{CG} C_{a\bf{n} b \bf{n}}C^a{}_{\bf{n}}{}^{b}{}_{\bf{n}}+\alpha_{CG} C_{abc\bf{n}}C^{abc}{}_{\bf{n}})\label{actionadm}
\end{equation}

where the projections of the Weyl tensor may be expressed as

\begin{eqnarray}
\perp{}^{\ms{(4)}}\!C_{\mathbf{n}a\mathbf{n}b}&=&\frac{1}{2}(2\gamma_{(a}^{\ c}\gamma_{b)}^{\ d}-\frac{1}{3}\gamma_{ab}\gamma^{cd})\Bigl[R_{cd}+K_{cd}K-\frac{1}{N}\left(\dot{K}_{cd}-\pounds_NK_{cd}-D_cD_dN\right)\Bigr],\label{electricWeyl} \\
\perp{}^{\ms{(4)}}\!C_{\mathbf{n}abc}&=&2(\gamma_{a}^{\ d}\gamma_{[b}^{\ [e}\gamma_{c]}^{\ f]}+\gamma_{a[b}\gamma_{c]}^{\ [e}\gamma^{\ f]d})D_fK_{ed},\label{magneticWeyl}
\end{eqnarray}
with  $\pounds_N$ denoting the Lie derivative along the shift vector $N^{a}$ on the spatial hypersurface. 
The explicit calculations of tensors (\ref{electricWeyl}) and (\ref{magneticWeyl}) can be found in the appendix of  \cite{Irakleidou:2014vla} .  $K_{ab}$ denotes extrinsic curvature of the  $\Sigma_t$  in terms of the time derivative of the metric $\gamma_{ab}$
 and the shift function $N_a$,
 \begin{equation} 
K_{ab}=\frac{1}{2N}\left(\dot{\gamma}_{ab}-D_{(a}N_{b)}\right).
\end{equation}

As it can be seen from the (\ref{actionadm}),  (\ref{electricWeyl}) and (\ref{magneticWeyl}) 
The  quantities that play a role of canonical coordinates and momenta make the canonical Lagrangian function of $\mathcal{L}(N,N^a,\gamma_{ab},\dot{\gamma}_{ab}, K_{ab}\dot{K}_{ab}, \lambda_{ab})$ with $\lambda_{ab}$   Lagrange multiplier which we will define shortly.

Unlike the Lagrangian formalism that can have arbitrary high order of time derivatives,  the Hamiltonian formalism is first order in time derivatives and it takes the form $\frac{df}{dt}=\{f,H\}$.
To reduce the CG action to first order in time derivatives 
 we use a trick connecting  $\gamma_{ab}$ and $K_{ab}$, two  independent variables,  via  
the constraint $\pounds_{\bf{n}}\gamma_{ab}-2 K_{ab}=0$, where $\pounds_{\bf{n}}$ is Lie derivative along the normal vector $n^a$. The constraint enters  the total Hamiltonian  multiplied by a Lagrange multiplier $\lambda_{ab}$.

The Lagrangian written in that form allows to determine the  constraints of the theory.
Since the action is independent of the time derivatives of lapse and shift, the canonical momenta conjugated to $N$ and $N^a$ respectively are primary constraints
\begin{eqnarray}
\Pi&\approx0&\\ 
\Pi_{a}&\approx0&.
\end{eqnarray} In which $\approx$ denotes a weak equality, as described in section 4. Canonical momenta conjugated to $\gamma_{ab}$ and $K_{ab}$ are respectively 
\begin{eqnarray}
\Pi_{\gamma}^{ab}&=&\frac{\partial \mathcal{L}}{\partial(\partial_t \gamma_{ab})}=\sqrt{\gamma}\lambda^{ab}\\
\Pi_{K}^{ab}&=&\frac{\partial \mathcal{L}}{\partial (\partial_c K_{ab})}=\sqrt{\gamma}(2\lambda C^{a}{}_{\bf{n}}{}^{b}{}_{\bf{n}})\label{pik}.
\end{eqnarray}
Inspecting (\ref{pik}) reveals it to be  traceless, which we have to take into account in the Lagrangian in the form of a constraint. We denote it by $\mathcal{P}$ and add it to Lagrangian by multiplication with the Lagrange multiplier $\lambda_P$.
The Lagrangian in the first order form  now reads
\begin{eqnarray}
L&=&\int_\Sigma \left(\Pi_K^{ab}\dot{K}_{ab}+\Pi_{\gamma}^{ab}\dot{\gamma}_{ab}-N\mathcal{H}_\ms{\perp}-N^a\mathcal{V}_a-\lambda_P\mathcal{P}\right)\nonumber\\
&\ &-\int_{\partial\Sigma}\ast\left[\Pi_K^{ab}D_bN-D_b\Pi_K^{ab}N+2N^c\left(\Pi_{\gamma}^{ab}\gamma_{bc}+\Pi_K^{ab}K_{bc}\right)\right],\label{Lagrangian3}\\
\mathcal{H}_{\ms{\perp}}&:=&-\omega_\gamma^{-1}\frac{\Pi_K^{ab}\Pi^K_{ab}}{2}+D_aD_b\Pi_K^{ab}+\Pi_K^{ab}\left(R_{ab}+K_{ab}K\right)-\omega_\gamma B_{abc}B^{abc}+2\Pi_{\gamma}^{ab}K_{ab},\\
\mathcal{V}_a&:=&\Pi_K^{bc}D_aK_{bc}-2D_b\left(\Pi_K^{bc}K_{ca}\right)-D_b\Pi_{\gamma}^{bc}\gamma_{ca},\\
\mathcal{P}&:=&\Pi_K^{ab}\gamma_{ab}.
\end{eqnarray}
in which $D$ is a covariant derivative on $\Sigma_t$  and * is  a contraction of the free vector index with the three dimensional volume form  $\omega_{\gamma}$ in the momenta.  The requirement that primary constraints are conserved in time leads to secondary constraints $\mathcal{H}_{\ms{\perp}}\equiv0$
 and $\mathcal{V}_{c}\equiv0$. 
 From the Lagrangian (\ref{Lagrangian3}), using the Legendre transforms and adding the primary constraints, we find the total Hamiltonian 
\begin{eqnarray}
H_T=\int_{\Sigma}\left(\lambda\Pi+\lambda^a\Pi_{a}+\lambda_P\mathcal{P}+N\mathcal{H}_\ms{\perp}+N^a\mathcal{V}_a\right)+\int_{\partial\Sigma}\left(\mathcal{Q}_\ms{\perp}+\mathcal{Q}_D\right),\label{eq:totalHamiltonian}
\end{eqnarray}
where surface contribution  $\int_{\partial\Sigma}\left(\mathcal{Q}_\ms{\perp}+\mathcal{Q}_D\right)$  is obtained such that it cancels the surface contribution  from  (\ref{Lagrangian3}), i.e.  $\mathcal{Q}_{\perp}$ and $\mathcal{Q}_D$  denote respectively: 
\begin{eqnarray}
\mathcal{Q}_\ms{\perp}&:=&\ast\left(\Pi_K^{ab}D_bN-D_b\Pi_K^{ab}N\right),\\
\mathcal{Q}_D&:=&2\ast N^c\left(\Pi_{\gamma}^{ab}\gamma_{bc}+\Pi_K^{ab}K_{bc}\right).
\end{eqnarray}
Following the procedure for consistency of the constraints below (\ref{eomphi}) we find that in order for $\mathcal{P}$ to be conserved in time one requires one more constraint $\mathcal{W}$
  \begin{equation}
\left\{H_T,\mathcal{P}\right\}\approx N\left(\Pi_K^{ab}K_{ab}+2\Pi_{\gamma}^{ab}\gamma_{ab}\right) =:N\mathcal{W},
\end{equation}
which leads to complete set of first class constraints,  which can be used to count the number of degrees of freedom of the theory. The initial set of 32 variables is reduced to 6 physical degrees of freedom that are associated with a massive and a partially massless graviton.

 
 \section{Castellani algorithm for Conformal gravity}
To find the canonical generators we have to determine primary first class constraints (PFCs). Via  calculation of Poisson bracket among the constraints we find PFCs  $\Pi, \Pi_a$ and $\frac{\mathcal{P}}{N}$ on which we apply the Castellani algorithm and obtain the gauge generators of the theory. To start the algorithm we take   $G_1$ to be  $\Pi, \Pi_a$ and $\frac{\mathcal{P}}{N}$ respectively, and  as an ansatz for right hand side for $G_1$ of $\Pi$ and $\frac{\mathcal{P}}{N}$,  we take the linear combination of the PFCs
\begin{equation}
\label{eq:PFCone}
PFC(x)=\int_{\Sigma} \,\,\,\left(\alpha(x, y)\Pi (y)+\beta^{a}(x,y){\Pi}_{a} (y)+\gamma(x, y)\mathcal{P}(y)\right).
\end{equation}
For the right hand side for $G_1$ of $\Pi_a$ we take similar ansatz such that the indices of the constraints and the coefficients are contracted.
Determination of the coefficients yields the generators of diffeomorphisms orthogonal to spatial hyper surface, i.e. transversal diffeomorphisms, $G_{\perp}$, spatial diffeomorphisms $G_D$, and  Weyl rescalings $G_{W}$

\begin{eqnarray}
\label{eq:g1}
G_{\ms{\perp}}[\epsilon, \dot{\epsilon}]&=& \int_\Sigma \! \left[\dot{\epsilon} \Pi+ \epsilon \left(\mathcal{H}_\ms{\perp}+\pounds_{N}\Pi+{\Pi}_{a} D^{a}N+D_{a}(\Pi^{a}N)+\frac{\lambda_{\mathcal{P}}}{N}\mathcal{P}\right)\right], \\
G_D[\epsilon^{a}, \dot{\epsilon}^{a}]&=& \int_\Sigma \left[ \dot{\epsilon}^{a} {\Pi}_{a}+ \epsilon^{a} \left(\mathcal{V}_{a}+\Pi D_{a}N+\pounds_{N}{\Pi}_{a}\right)\right],\label{eq:g2}\\
G_{W}[w, \dot{w}]&=& \int_\Sigma \! \left[ \frac{\dot{w}}{N}\mathcal{P} (x)+ w \left( \mathcal{W}+N\Pi+\pounds_{N}\frac{\mathcal{P}}{N}\right)\right].\label{eq:g3}
\end{eqnarray}
In which the $\epsilon s$ from (\ref{defg}) are denoted with $\epsilon, \epsilon^a$ and $w$.
If the generators (\ref{eq:g1}), (\ref{eq:g2}) and (\ref{eq:g3}) are functionally differentiable we are able to determine the canonical charges of the theory. This is in general done in the following way. Let us consider the theory with the fields $\{\phi_i\}$, with a gauge transformation  $\xi$  playing the role of $\epsilon$ in (\ref{defg}). 
Gauge transformation generated by $\xi$ is
\begin{equation}
G[\xi,\phi]=\int_{\Sigma}d^nx\mathcal{G}[\xi,\phi].
\end{equation}
For the deformation of $G[\xi,\phi]$ under general variation  
\begin{equation}
\delta G[\xi,\phi]=\int_{\Sigma}d^nx\frac{\delta \mathcal{G}}{\delta \phi_i} \delta\phi_i+\int_{\partial\Sigma}d^{n-1}x B[\xi,\phi,\delta\phi]. \label{integrability}
\end{equation}
to be well defined $B$ has to be zero, in which case G is differentiable. 
Therefore, we have to choose the boundary term  such that B is total variation 
\begin{eqnarray}
B[\xi,\phi,\delta\phi]&=&-\delta\Gamma[\xi,\phi]\\
Q[\xi]&=&\Gamma[\xi].
\end{eqnarray}
Where we denoted the term that has to be added to the generator in order for its total variation to be well defined with $\Gamma$, which corresponds to the charge of the theory.

\section{Canonical charges}

Let us remember the expression  (\ref{eq:totalHamiltonian}). It turned out that it is well defined with the boundary terms
\begin{eqnarray}
\mathcal{Q}_\ms{\perp}&:=&\ast\left(\Pi_K^{ab}D_bN-D_b\Pi_K^{ab}N\right),\\
\mathcal{Q}_D&:=&2\ast N^c\left(\Pi_{\gamma}^{ab}\gamma_{bc}+\Pi_K^{ab}K_{bc}\right).
\end{eqnarray}
Analogously,
for the boundary term in (\ref{integrability}) to vanish one has to add suitable boundary terms to the variation of the generators (\ref{eq:g1}), (\ref{eq:g2}) and (\ref{eq:g3}) . This leads to a new set of improved generators.
Varying the  generator (\ref{eq:g3}) one notices that in the remaining boundary integral only the constraint $\mathcal {P}$ appears, making the boundary part vanish on shell. Therefore the boundary contribution for that generator is zero from which we conclude that the corresponding charge, the Weyl charge, is zero. The boundary term that must be added to the generator of spatial diffeomorphisms   (\ref{eq:g1}) corresponds to the expression we are already familiar with

\begin{equation}
Q_D[\epsilon]=2 \int_{\partial\Sigma}\ast\left(\Pi_{\gamma}^{ab}\gamma_{bc}+\Pi^{ab}_{K}K_{bc}\right)
\end{equation}
which therefore corresponds to the spatial diffeomorphism charge. In order for the generator (\ref{eq:g1}) to yield the vanishing boundary term, one has to add 
\begin{equation}
Q_\ms{\perp}[\epsilon_\ms{\perp}]=\int_{\partial\Sigma}\ast\left(\Pi_K^{ab}D_b\epsilon_\ms{\perp}-D_b\Pi_K^{ab}\epsilon_\ms{\perp}\right).\label{eq:perpcharge}
\end{equation}
which corresponds to the charge of transversal diffeomorphisms. 
One can show that the charges are finite by explicitly inserting the Fefferman-Graham expansion and using the boundary conditions defined in \cite{Grumiller:2013mxa}. However, there is an alternative way to prove this. Rescaling the metric with arbitrary scale factor $\Omega$ and applying the appropriate rescaling on each of the tensorial quantities, one can show  charges are finite for even when the boundary conditions are relaxed \cite{Irakleidou:2014vla}.


There are multiple ways of showing that the charges defined above are conserved. For instance, it can be established by showing that the charges are equivalent to the Noether charges found in \cite{Grumiller:2013mxa}. This was carried out in \cite{Irakleidou:2014vla}.

\section{Asymptotic Symmetry Algebra}

The boundary conditions imposed on conformal gravity such that it has a well defined variational principle and finite response function determine the boundary conditions preserving transformations and subsequently the Asymptotic Symmetry Algebra (ASA) \cite{Grumiller:2013mxa}. 
They are generated by differomorphisms $\pounds_\xi$ with vector field $\xi$ and Weyl transformations $\delta_{\omega}$ that are generated  by scalar field $\omega$ \cite{Grumiller:2013mxa}.  The Lie algebra of the boundary condition preserving gauge transformation is isomorphic to the Dirac algebra of the charges. 
They impose conditions on the terms in the Fefferman-Graham expansion of the  metric to leading and next-to-leading order in holographic coordinate up to Weyl rescalings of the metric.

Taking the boundary conditions 
\begin{equation}
ds^2=\frac{e^{2\omega}\ell^2}{\rho^2}\left(d\rho^2+\gamma_{ab}dx^adx^b\right),\label{eq:BC4}
\end{equation}
the boundary condition preserving gauge transformations requires
\begin{equation}
\pounds_\xi \frac{\ell^2}{\rho^2}\overline{g}_{\mu\nu}=2\lambda\frac{\ell^2}{\rho^2}\overline{g}_{\mu\nu},\label{bconds}
\end{equation}
with $\overline{g}_{\rho\rho}=1$, $\overline{g}_{ab}=\gamma_{ab}$, $\bar{g}_{a \rho} = 0$, and $\ell$ the AdS length scale.  

Expanding the functions appearing in (\ref{eq:BC4}) in terms of the holographic coordinate $\rho$, one finds that the leading orders of  $\xi(\rho)$ and $\lambda$  have to vanish, as well as of the first order of  $\xi^a$, where we have assumed that the expansion is homogeneous in $\rho$, i.e. there are no logarithmic terms in the expansion of the holographic coordinate. The condition on the leading order term in the expansion of the  metric is 

\begin{equation}
\pounds_{\xi^a_{\LO}}\gamma^{\LO}_{bc}=\frac{2}{3}\mathcal{D}_d\xi^d_{\LO}\gamma^{\LO}_{bc},\label{eq:LOCKV}
\end{equation}
so the $\xi^{a}_{(0)}$ are conformal Killing vectors (CKVs) of the boundary metric $\gamma^{(0)}_{ab}$. If the boundary metric is conformally flat (i.e., has vanishing Cotton tensor) then there are 10 such CKVs.
The next-to-leading order term $\gamma^{(1)}_{ab}$ is part of the boundary data, and the diffeomorphisms that preserve the boundary conditions must satisfy
\begin{equation}
\pounds_{\xi^a_{(0)}}\gamma_{bc}^{(1)}=\frac{1}{3}\mathcal{D}_{d}\xi^{d}\gamma^{(1)}_{bc}-4\omega^{(1)}\gamma_{bc}^{(0)}. \label{eqlo}
\end{equation}
This identifies a subset of the diffeomorphisms that satisfy (9.3). A classification of the possible ASAs can be found in \cite{refcl}.
Here we consider two examples where non-trivial boundary data $\gamma^{(1)}_{ab}$ identifies a sub-algebra of the full conformal algebra as the ASA.

The first example is the solution of the Bach equation that  simultaneously has flat Ricci tensor. Its line element can be written as

\begin{equation} 
ds^2=dr^2-(1+r x) dt^2+2 r x dtdx + dx^2+(1+r x) dy^2.
\end{equation}

Substituting $r \rightarrow \rho$ we can immediately see that the line element is 
already conformal to the Fefferman-Graham form (9.1), and we can read off the boundary data

\begin{eqnarray}
\gamma^{(0)}_{ab}&=&diag(-1,1,1)\\
\gamma^{\FO}_{ab}&=&\left(\begin{array}{c c c} x & 0& x \\ 0& 0 &0 \\ x& 0 & x\end{array}\right).
\end{eqnarray}

The boundary metric admits 10 CKVs, and the condition (\ref{eqlo}) identifies the following three vectors as generators of the ASA


 \begin{eqnarray}
 \xi^{\LO a}_{1}&=&(0,0,1),\\
\xi^{\LO a}_{2}&=&(t-y,x,-t+y), \\
\xi^{\LO a}_{3}&=&(1,0,0).\\
 \end{eqnarray}
 The second generator is a linear combination of the CKVs that generate dilatations and boosts in the t-y plane. Defining $\xi^{\LO a}_{+}=\xi^{\LO a}_{1}+\xi^{\LO i}_{3}$, $\xi^{\LO a}_{-}=\xi^{\LO a}_{1}-\xi^{\LO a}_{3}$ and $\chi^{(0) a}=\frac{1}{2}\xi^{\LO a}_2$ we obtain the ASA $[\chi^{(0) a},\xi^{\LO a}_{-}]=- \xi^{\LO a}_{-}$.
 

Next we consider  the Mannheim-Kazanas-Riegert (MKR) solution. 
This is the most general spherically symmetric solution of conformal gravity, analogous to the Schwarzschild solution of Einstein gravity.
The MKR line element reads
\begin{equation}
ds^2=-k(r)dt^2+\frac{dr^2}{k(r)}+r^2d\Omega_{S^{2}}^2
\end{equation}
with
\begin{equation}
k(r)=\sqrt{1-12 a M}-\frac{2M}{r}-\frac{\Lambda r^2}{3}+2 a r.
\end{equation}
Rewriting this metric in the Fefferman-Graham form (\ref{eq:BC4}) gives the boundary data

\begin{eqnarray} 
\gamma^{\LO}_{ab}&=&diag(-1,\ell^2,\ell^2\sin^2\theta), \\
\gamma^{\FO}_{ab}&=&\left(\begin{array}{c c c}0 & 0& 0 \\ 0& -2a\ell^3 &0 \\ 0& 0 & -2a\ell^3\sin^2\theta\end{array}\right).
\end{eqnarray} 
As before, the boundary metric admits 10 CKVs. Four of these satisfy (\ref{eqlo})

 \begin{eqnarray}
 \xi^{\LO a}_{1}&=&(0,0,1),\\
\xi^{\LO a}_{2}&=&(0, \sin (\phi), \cot (\theta)\cos (\phi)),\\
\xi^{\LO a}_{3}&=&(0, -\cos (\phi), \cot (\theta)\sin (\phi)), \\
\xi^{\LO a}_{4}&=&(1,0,0).\\
 \end{eqnarray}
These generators give the ASA $\mathbb{R} \oplus \mathcal{SO}(3)$.

\section{Conclusion and Outlook}

In the talk we have heard about the canonical analysis of one of the higher derivative theories, conformal gravity. The analysis consisted of construction of canonical generators corresponding to gauge symmetries, usage of the Castellani alogrithm, applying differentiability
 conditions and obtaining the canonical charges which are integrable, finite and time conserved. The charge associated to Weyl rescalings is zero making this a trivial gauge transformation which is in accordance with analysis in \cite{Campigotto:2014waa} and differs from the case in three dimensions, where the Weyl invariant higher derivative action, Cern-Simons  action, has vanishing Weyl charge in case when the Weyl factor is not allowed to vary freely. When the Weyl factor is allowed to vary, the associated charge is not vanishing \cite{Afshar:2011qw}. 

The charges generate asymptotic symmetry algebra that under particular conditions - with particular choice of the first term in the Fefferman-Graham expansion of the metric - leads to full conformal algebra which is  further restricted by the conditions from the next-to-leading order term in the expansion of the metric. 
We have shown two such examples. In the first example  $\mathcal{SO}(2,3)$ conformal algebra is restricted to sub algebra formed by the three  Killing vectors while on the example of Mannheim-Kazanas-Riegert solution that restriction leads from   $\mathcal{SO}(2,3)$ algebra in the leading order to $R\oplus\mathcal{SO}(3)$ in the next to leading order.

\section{Acknowledgements}

I would like to thank Daniel Grumiller, Robert McNees, Florian Preis and Maria Irakleidou for the careful reading of the text  and for the discussions. Some of the calculations were done using the {\it Mathematica} packages {\it xAct} \cite{xAct}  and {\it RGTC} \cite{RGTC}. This work was supported by the START project Y 435-N16 and by the project I 952-N16 of the Austrian
Science Fund (FWF).

\bibliography{bibliothek}

\end{document}